\documentclass[12pt,aps,tightenlines,notitlepage]{revtex4-2}
\usepackage{amsmath}
\usepackage{amssymb}
\usepackage{bm}
\usepackage[mathscr]{eucal}
\usepackage{graphicx}
\usepackage{psfrag}
\usepackage{subfigure}
\usepackage{color}
\usepackage{wasysym}
\include{graphix}
\usepackage{framed}
\usepackage{enumitem}%
\usepackage{lipsum}
\usepackage{float}
\usepackage{stmaryrd}
\usepackage{MnSymbol}
\definecolor{orange}{rgb}{1,0.5,0}
\usepackage[normalem]{ulem}

\usepackage{tikz}

\usetikzlibrary{patterns}
\usetikzlibrary{shapes}
\def\normaltwo{\x,{0.7+0.8/exp(((\x-6)^2)/2)}}

\graphicspath{{figs/}}

\newtheorem{theorem}{Theorem}[section]

\newenvironment{remark}[1][Remark]{\begin{trivlist}
\item[\hskip \labelsep {\bfseries #1}]}{\end{trivlist}}

\newcommand{\myqed}{\nobreak \ifvmode \relax \else
      \ifdim\lastskip<1.5em \hskip-\lastskip
      \hskip1.5em plus0em minus0.5em \fi \nobreak
      \vrule height0.75em width0.5em depth0.25em\fi}

\newcommand{\mathd}{\mathrm{d}}
\newcommand{\mathe}{\mathrm{e}}

\newcommand{\myRe}{\mathrm{Re}}
\newcommand{\myWe}{\mathrm{We}}
\newcommand{\hbl}{h_{bl}}

\newcommand{\thetadyn}{\vartheta_{a}}
\newcommand{\surften}{\sigma}

\newcommand{\rr}{\mathcal{R}_{max}}

\newcommand{\myconst}{k}
\newcommand{\rmax}{R_{max}}

\newcommand{\newconst}{\chi}
\newcommand{\newR}{\tilde{R}}
\newcommand{\newt}{\tilde{t}}
\newcommand{\newT}{\tilde{T}}

\newcommand{\dittotikz}{%
    \tikz{
        \draw [line width=0.12ex] (-0.2ex,0) -- +(0,0.8ex)
            (0.2ex,0) -- +(0,0.8ex);
        \draw [line width=0.08ex] (-0.6ex,0.4ex) -- +(-1.5em,0)
            (0.6ex,0.4ex) -- +(1.5em,0);
    }%
}

\begin{document}

%+Title
\title{Bounds on the Spreading Radius in Droplet Impact: The Viscous Case}
%\author{Alidad Amirfazli$^1$, Miguel Bustamante$^2$, Yating Hu$^1$, Lennon \'O N\'araigh$^2$, Anyone Else}
\author{Lennon \'O N\'araigh$^1$, Miguel D. Bustamante$^1$ }
\email{onaraigh@maths.ucd.ie}
%\address{$^1\,$Department of Mechanical Engineering, York University, Toronto, Ontario M3J 1P3, Canada\\
%$^2\,$School of Mathematics and Statistics, University College Dublin, Belfield, Dublin 4, Ireland}
\address{$^1\,$School of Mathematics and Statistics, University College Dublin, Belfield, Dublin 4, Ireland}
\date{\today}

\begin{abstract}
We consider the problem of droplet impact and droplet spreading on a smooth surface in the case of a viscous Newtonian fluid.  We revisit the concept of the rim-lamella model, in which the droplet spreading is described by a system of ordinary differential equations (ODEs).  We show that these models contain a singularity which needs to be regularized to produce smooth solutions, and we explore the different regularization techniques in detail.  We adopt one particular technique for further investigation, and we use differential inequalities  to derive upper and lower bounds for the maximum spreading radius $\rr$.  Without having to resort to dimensional analysis or scaling arguments, our bounds confirm the leading-order behaviour $\rr \sim \myRe^{1/5}$, as well as a correction to the leading-order behaviour involving the combination $\myWe^{-1/2}\myRe^{2/5}$, in agreement with  the experimental literature on droplet spread.    Our bounds are also consistent with full numerical solutions of the ODEs, as well as with energy-budget calculations, once head loss is accounted for in the latter.  
\end{abstract}

\maketitle

\section{Introduction}
\label{sec:intro}

Understanding the physics of droplet on impact on a solid wall is relevant many practical applications, e.g. inkjet printing~\cite{yarin2006drop}, cooling~\cite{yarin2006drop,valluri2015}, and crop spraying~\cite{yarin2006drop,Moghtadernejad2020}, as well as environmental applications~\cite{tsimplis1989wave}.  Indeed, the range of applications is burgeoning, as evidenced by recent reviews articles~\cite{mohammad2023physics,wang2023droplet}.  Different droplet-impact scenarios can occur, depending on the droplet's Weber number ($\myWe$) and Reynolds number ($\myRe$).  For instance, there is a splash parameter $K=\myWe\sqrt{\myRe}$, which determines a threshold above which
splash occurs~\cite{mundo1995droplet,josserand2016drop}.  The threshold value is not universal~\cite{marengo2011drop}, and different experiments have produced different values, a summary of which is provided in~\cite{moreira2010advances}.  Nevertheless, the threshold is of the order of $K=10^3-10^4$.

Just below this threshold, and typically for $\myWe \geq 10^2$ and $\myRe \geq 10^3$~\cite{de2010thickness}, there is `rim-lamella' regime,
in which the droplet flattens and spreads into an axisymmetric structure involving a lamella, with a thicker rim forming at the extremity. A key parameter which characterizes the droplet impact in this regime is the maximum spreading radius, 
denoted here by $\rr$, and is governed by  $\myWe$ and $\myRe$. For consistency with a previous work~\cite{amirfazli2023bounds}, we use the definitions
\begin{equation}
\myWe=\frac{\rho U_0^2 R_0}{\sigma},\qquad \myRe=\frac{\rho U_0 R_0}{\mu},
\label{eq:maindefs}
\end{equation}
where $\rho$ is the liquid density, $\mu$ is the liquid viscosity, and $\sigma$ the surface tension. Also, $U_0$ is the
droplet's speed prior to impact and $R_0$ is droplet radius prior to impact.  Of interest in this regime is the maximum spreading, $\rr$, which depends on $\myWe$ and $\myRe$.

There is no exact formula for $\rr(\myWe, \myRe)$.  In the limit of very large Reynolds number, boundary-layer theory and dimensional analysis tell us that $\rr \sim \myRe^{1/5}$.  A wide range of experiments have been performed which confirm this scaling, and which further reveal the contribution of the Weber number.  These results give $D_{max}/D_0=0.87 \myRe_{D}^{1/5}-0.40\myRe_D^{2/5}\myWe_D^{-1/2}$.  Here, the subscript $D$ denotes non-dimensionalization on the droplet diameter, $D_0=2R_0$.  Correspondingly, $D_{max}$ is the maximum spreading diameter.  In terms of non-dimensionalization based on $R_0$, the established correlation is:
\begin{equation}
\frac{\rr}{R_0}=1.00 \myRe^{1/5}-0.34 \myRe^{2/5}\myWe^{-1/2},
\label{eq:correlation}
\end{equation}
a result first obtained by Roisman \textit{et al.}~\cite{roisman2002normal} by collapsing a wide range of experimental data to a single curve.  

Theoretical modelling work has already been done, to understand the dependence of $\rr$ on $\myWe$ and $\myRe$.  Such work has been based on energy balance or momentum balance (rim-lamella models).  We describe this modelling work below in the literature review.  We mention in particular the rim-lamella model due  by Roisman \textit{et al.}~\cite{roisman2002normal}, which was later extended by Eggers \textit{et al.}~\cite{eggers2010drop}, to account for the viscous boundary layer.  Further development of the model has been done by Gordillo \textit{et al.}~\cite{gordillo2019theory}, who used perturbation theory to fill in the details of the viscous flow structure inside the lamella.

\subsection{Aims of the Paper}

We introduce a modified rim-lamella model similar to the one by Eggers \textit{et al.}~\cite{eggers2010drop} and Gordillo \textit{et al.}~\cite{gordillo2019theory}.  In doing so, we have three main aims.  First,  we  develop the model rigorously from first principles.  We uncover a term which makes the model ill-posed, in the sense that a smooth solution can not always be found.  We review the different regularization techniques that have been employed (implicitly) in the literature.  We state carefully the regularization technique employed in this work.  This enables us to fuflil our second aim, which is to conduct an analysis of the rim-lamella model, based  on the theory of differential inequalities.  We show the existence of rigorous bounds:
\begin{equation}
\myconst_1 \myRe^{1/5}-\myconst_2\sqrt{(1-\cos\thetadyn)} \myWe^{-1/2}\myRe^{2/5}\leq 
\frac{\rr}{R_0} \leq \myconst_1 \myRe^{1/5},
\label{eq:bounds}
\end{equation}
where $\myconst_1$ and $\myconst_2$ are constants (independent of $\myRe$ and $\myWe$), and $\thetadyn$ is the advancing contact angle.  
A final  aim of the paper is to use the regularized rim-lamella model to provide an independent means of validating the energy-balance approach to predicting the maximum spreading radius.   In particular, we will use  the rim-lamella model to establish an independent estimate of the `head loss', a key term for producing an accurate energy-budget model.

\subsection{Literature Review}

There are two main theoretical methods for estimating the maximum spreading radius based on $\myRe$ and $\myWe$.  These are the energy-budget method and rim-lamella models.
In the energy-budget method (e.g.~\cite{chandra1991collision}), the pre-impact energy of the droplet is related to the droplet energy at maximum spreading, which is assumed in a first approximation to be entirely due to surface energy.    By treating the droplet at maximum spreading as having a pancake-like structure,
a simple closed expression can be found for $\rr$ as a function of $\myWe$. Viscous dissipation can be
incorporated into the energy budget by explicitly modelling the boundary layer in the lamella~\cite{pasandideh1996capillary}.  However, it is only by incorporating the concept of `head loss' into the energy  budget that an accurate agreement between the theory, experiments, and direct numerical simulation be obtained~\cite{naraigh2023analysis}.  In this context, `head loss' means the energy conversion of initial kinetic energy into internal fluid motion in the rim-lamella structure~\cite{villermaux2011drop}, and is added into the energy budget as an extra term.  By assuming that the the head loss is equal to one half the initial kinetic energy, excellent agreement between the energy-budget model and the semi-empirical correlation~\eqref{eq:correlation} is obtained~\cite{wildeman2016spreading}.  %Therefore, another aim of this work is to justify the head-loss assumption using the rim-lamella model as a second independent approach.

The rim-lamella model was introduced by Roisman \textit{et al.}~\cite{roisman2002normal} for the inviscid case.  Roisman \textit{et al.} subsequently highlighted the importance of the viscous boundary layer and its effect on droplet spreading~\cite{roisman2009inertia}.  Eggers \textit{et al.} combined these descriptions and introduced a rim-lamella model for the viscous case, which exhibits the leading-order scaling behavior $\rr/R_0 \sim \myRe^{1/5}$ for $\myRe\gg 1 $.   This basic rim-lamella model
has been greatly extended in recent papers by Gordillo \textit{et al.}~\cite{gordillo2019theory,garcia2020inclined}.    In these works, a detailed expression for the flow in the lamella is developed by expanding the momentum-balance equation in powers of $\myRe^{-1}$, and solving the resulting hierarchy of equations by the method of characteristics.

These rim-lamella models have some crucial advantages.  Although some fitting of the paramters associated with the boundary layer is required to obtain agreement with the experimental results, there is no need to introduce missing physics, such as `head loss'.  All the physics is contained in the model.  Also, as the rim-lamella is a system of ordinary differential equations for the lamella extent, the lamella height, and the rim volume, the model gives some insight into the dynamics leading up to maximum spreading, as well the maximum spreading itself.  The model gives further insight into the evolution after maximum spreading, where the rim retracts in a phenomenon that is very similar to the Taylor--Culick retraction of a liquid sheet.  Hence, the rim-lamella models combine simplicity (a system of three ODEs can be solved within minutes if not seconds on a desktop computer), with insights into the dynamics of spreading.

The relatively simple model due to Eggers \textit{et al.} is particularly attractive because it encapsulates the basic physics of spreading and retraction.  
For engineering applications, such a simple model that still provides a reasonably accurate representation of the droplet spreading and retraction process is desirable, e.g. for control of droplet spreading~\cite{hu2024application}, modelling droplet spreading on rough surfaces~\cite{tang2017dynamics}, or spreading during evaporation~\cite{lyu2021dynamics}.
The same approach may be useful when control of droplet spreading is modelled numerically, using optimal control theory.  Recent numerical works use optimal control theory both to move sessile droplets, and to control their shape~\cite{bonart2020optimal,shankar2022optimal}.  Having a robust model for these purposes is essential, as numerical algorithms for optimal control (e.g. those based on steepest-descent) require the model to be solved over many iterations and over many different parameter configurations (which are not known \textit{a priori}), until convergence is reached~\cite{biral2016notes}.   

Regardless of whether the approach of Eggers \textit{et al.} or Gordillo \textit{et al.} is followed, the resulting rim-lamella model contains a term which makes it ill-posed.  Consequently, a regularization of the model must be carried out, which to date has only been done implicitly.  We make the regularization explicit in this work, which enhances our understanding of this class of models.  In doing so, we introduce a simple and robust rim-lamella model tailored for the engineering applications alluded to above, but which also facilitates rigorous analysis, based on the theory of differential inequalities.

The use of differential inequalities in Fluid Mechanics is particularly fruitful, as it yields rigorous results for highly complex systems of systems of equations.  To date, the theory  has been used to put constraints on the critical Reynolds number in nonlinear instability of channel flow~\cite{doering1995applied}, on the expected regularity of the solutions of the Navier--Stokes equations~\cite{doering1995applied}, and on the location of the eigenvalues for the stability of one-dimensional shear flows~\cite{howard1961note}.  The theory has also been used to quantify mixing efficiency~\cite{thiffeault2004bound}, and to identify the conditions under which thin films will rupture~\cite{naraigh2010nonlinear}.  We have also used the theory in a previous work~\cite{amirfazli2023bounds} to put bounds on the spreading radius in droplet impact in the inviscid case ($\myRe=\infty$).  In that case, we showed: 
\begin{equation}
\frac{\rr}{R_0}=\myWe\, f(\myRe),\qquad \myWe\gg 1,
\end{equation}
where $f(\myRe)$ is a bounded function, $f(\myWe)\leq \text{Const.}$  
Hence, in the present paper, we extend the work in Reference~\cite{amirfazli2023bounds} to account for viscosity, with a view to developing a relatively simple model that can   potentially be extended to other applications involving more complex physics.

\subsection{Plan of the Paper}

In Section~\ref{sec:model} we develop a rim-lamella model from first principles.  We outline how the model can develop a singularity, and describe how the singularity can be regularized.
In this way, we develop a simple rim-lamella model amenable to theoretical analysis.   In Section~\ref{sec:theory} we present some preliminary theoretical based on differential inequalities.  We continue this work in Section~\ref{sec:bounds} where we develop  upper and lower bounds on the spreading radius and establish the result~\eqref{eq:bounds}.  We evaluate the bounds in Section~\ref{sec:evaluation}.  Specifically, we investigate the sharpness of the bounds with respect to numerical solutions of the rim-lamella model.  We also compare the results of the rim-lamella model to the energy-budget model by Wildeman et al.~\cite{wildeman2016spreading}.   Concluding remarks are presented in Section~\ref{sec:conc}.

\section{Mathematical Model}
\label{sec:model}

In this section we introduce a rim-lamella model which takes account of the viscous boundary layer which forms after droplet impact.  Our version builds on the model by Eggers \textit{et al.}~\cite{eggers2010drop} and addresses the singularity which occurs in that model for certain realistic initial conditions, we which describe in detail.  At the same time, the model is a simplification with respect that of Gordillo \textit{et al.}~\cite{gordillo2019theory}, which is desirable for the intended applications mentioned in the introduction.

%The inviscid rim-lamella model was first introduced by Roisman \textit{et al.}~\cite{roisman2002normal}.  A viscous version including a description of the boundary layer was introduced by Eggers \textit{et al.}~\cite{eggers2010drop}.  Our version builds on the model by Eggers but is mathematically well posed, and avoids a  singularity which occurs in that model for certain (realistic) initial conditions. 

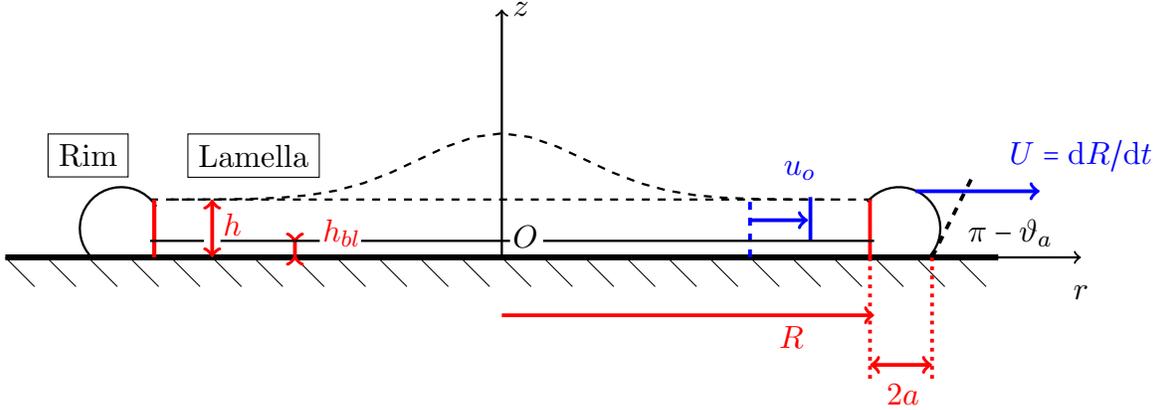
\begin{figure}[htb]
\centering
\begin{tikzpicture}[scale=1.1, transform shape]
\draw[-,black,line width=0.8mm] (0,0) -- (12,0);
\foreach \x in {-1,...,22}
\draw (0.5+0.5*\x,0) -- (0.5+0.5*\x+0.5*0.707,-0.5*0.707);
\draw[->,black,line width=0.3mm] (6,0) -- (13,0);
\draw (13, -0.2) node[below] {$r$};
\draw[->,black,line width=0.3mm] (6,0) -- (6,3);
\draw (6, 3) node[right] {$z$};
\draw (6 ,0.25) node[right] {$O$};
%
% % Top Gaussian curve:
\draw[black,dashed,line width=0.3mm,domain=2:10] plot (\normaltwo) node[right] {};
%
% % Top Solid Line:
\draw[black,dashed,line width=0.3mm] (1.75,0.7) -- (10.5,0.7);
%
% % Rims:
\draw [black,line width=0.3mm,domain=-45:140] plot ({10.8+0.5*cos(\x)}, {0.35+0.5*sin(\x)});
\draw [black,line width=0.3mm,domain=32:228] plot ({1.4+0.5*cos(\x)}, {0.35+0.5*sin(\x)});
%
% % Marker for height h:
\draw [<->,red,line width=0.5mm] (2.5,0) -- (2.5,0.7);
\draw (2.5, 0.4) node[right] {${\color{red}{h}}$}; 
%
% % Marker for boundary-layer height h_bl:
%
\draw [>-<,red,line width=0.5mm] (3.5,-0.1) -- (3.5,0.3);
\draw (3.7, 0.3) node[right] {${\color{red}{\hbl}}$}; 
%
%
% % Label for R, lamella radius:
\draw [->,red,line width=0.5mm] (6,-0.7) -- (10.5,-0.7);
%\draw (10.5,-0.7) node[below] {${\color{red}{R}}$}; 
\draw (9.5,-0.7) node[below] {${\color{red}{R}}$}; 
\draw [-,red,line width=0.5mm] (10.45,0) -- (10.45,0.7);
\draw [-,red,line width=0.55mm] (1.8,0) -- (1.8,0.7);
%
%
% % Droplet footprint thickness:
%\draw [dotted,red,line width=0.5mm] (10.45,0) -- (10.45,-0.45);
%\draw [dotted,red,line width=0.5mm] (10.45,-1.25) -- (10.45,-1.5);
\draw [dotted,red,line width=0.5mm] (10.45,0) -- (10.45,-1.5);
\draw [dotted,red,line width=0.5mm] (11.2,0) -- (11.2,-1.5);
\draw [<->,red,line width=0.5mm] (10.45,-1.3) -- (11.2,-1.3);
\draw (10.85, -1.4) node[below] {${\color{red}{2a}}$}; 
%
% % Rim and lamella labels:
\draw (3, 1.5) node[rectangle,draw,below] {Lamella};
\draw (1, 1.5) node[rectangle,draw,below] {Rim};
\draw [black,dashed,line width=0.5mm] (11.2,0) -- (11.7,1);
\draw (11.5, 0.3) node[right] {$\pi-\thetadyn$};
\draw [->,blue,line width=0.5mm] (11,0.8) -- (12.5,0.8);
\draw (13,0.9) node[above] {{\textcolor{blue}{$U=\mathd R/\mathd t$}}};
%
% % Mean Flow:
\draw [->,blue,line width=0.5mm] (7.5+1.5,0.45) -- (8.2+1.5,0.45);
\draw [-,blue,line width=0.5mm] (9.73,0.2) -- (9.73,0.73);
%\draw [-,blue,line width=0.5mm] (9,0) -- (9.73,0.34);
\draw [blue,dashed,line width=0.5mm] (9,0) -- (9,0.7);
\draw (9.6,0.8) node[above] {{\textcolor{blue}{$u_o$}}};
%
%
% Hyperbolic flow:
%
%\draw[color=blue,dashed,line width=0.3mm,domain=7:12] plot (\hyperplus) node[right] {};
%
% Bottom line - the flat boundary layer: 
\draw[color=black,line width=0.3mm] (1.75,0.2) -- (2.4,0.2);
\draw[color=black,line width=0.3mm] (2.6,0.2) -- (3.8,0.2); 
\draw[color=black,line width=0.3mm] (4.3,0.2) -- (6.1,0.2);
\draw[color=black,line width=0.3mm] (6.5,0.2) -- (10.5,0.2);
%
% % Curved Boundary Layer (redunant):
%\draw[color=black,line width=0.3mm,domain=7.5:10.6] plot (\blone) node[right] {};
%\draw[color=black,line width=0.3mm,domain=11.6:11.8] plot (\blone) node[right] {};
%
%\draw[color=black,line width=0.3mm,domain=0.4:4.5] plot (\bltwo) node[right] {};
%\draw[color=black,line width=0.3mm,domain=0.4:2.5] plot (\bltwo) node[right] {};
%\draw[color=black,line width=0.3mm,domain=3:4.5] plot (\bltwo) node[right] {};
%
\end{tikzpicture}
\caption{Schematic diagram showing the cross-section of an axisymmetric rim-lamella structure.  Dashed curved line:  true lamella height $h(r,t)=(t+t_0)^{-2}g[(r/(t+t_0)]+h_{PI}(t)$, as given by Equation~\eqref{eq:ht1}.    Dashed straight line:  the remote asymptotic approximation $h(r,t)=h_{init}[(\tau+t_0)/(t+t_0)]^2+h_{PI}(t)$.}
\label{fig:schematic}
\end{figure}

The schematic diagram for the model is shown in Figure~\ref{fig:schematic}.  The flow far from the substrate  is denoted by $u_o$ (the outer flow).  Following previous works, this is assumed to be a hyperbolic flow, and is given by the expression
$u_o=r/(t+t_0)$, where $t_0$ is a model parameter which describes the onset time of the flow, prior to the formation of the rim-lamella structure.  In contrast to the inviscid case, a boundary layer forms close to the substrate, in which the fluid velocity transitions from $u_o$ to zero, across a distance $\hbl$, the boundary-layer thickness.  From boundary-layer 
theory~\cite{white2011fluid}, the boundary layer thickness scales as $\hbl\propto \sqrt{\nu r/u_o}$, where $\nu=\mu/\rho$ is the kinematic viscosity of the liquid.  Given the functional form of $u_o$, the $r$-dependence cancels out, to give $\hbl\propto \sqrt{\nu (t+t_0)}$.  We allow for an additional degree of freedom in the problem by taking $\hbl\propto\sqrt{\nu (t+t_1)}$, where $t_1$ describes the time of formation of the boundary layer.  

We refer again to boundary-layer theory~\cite{white2011fluid}, and describe the variation of the flow in the vertical direction with a transition function, such that:
\begin{equation}
u(r,z,t)=u_o(r,t)F(z),
\label{eq:utotal}
\end{equation}
where $F(z=0)=0$ and $F(z)=1$ for $z\geq \hbl$. 
We recall three standard models for the transition function $F(z)$:
\begin{subequations}
\begin{eqnarray}
\text{Step function:}&\phantom{=}& F(z)=\begin{cases} 0,& z<\hbl,\\
                                                       1,& z>\hbl,\end{cases},\\
\text{Linear function:}&\phantom{=}& F(z)=\begin{cases} \frac{z}{\hbl},& z<\hbl,\\
                                                       1,& z>\hbl,\end{cases},\\
\text{Quadratic function:}&\phantom{=}& F(z)=\begin{cases} 2\left(\frac{z}{\hbl}\right)-\frac{z^2}{\hbl^2},& z<\hbl,\\
                                                       1,& z>\hbl,\end{cases}.																																																						
\end{eqnarray}%
\label{eq:transition}%
\end{subequations}%
For $h>\hbl$ this gives:
\begin{equation}
\int_0^h F(z)\,\mathd z=h-\tfrac{1}{n}\hbl,
\label{eq:nval}
\end{equation}
where $n=1$ for the step function, $n=2$ for the linear function, and $n=3$ for the quadratic function.
We use the incompressibility condition $r^{-1}\partial_r(ru)+\partial_z w=0$ to obtain an expression for $w(r,z,t)$:
\begin{equation}
w(r,z,t)=-\frac{2}{t+t_0}\int_0^z F(z)\mathd z.
\end{equation}
% hence $w(r,z,t)=-2/(t+t_0)$ if $z>\hbl$ and $w(r,z,t)=0$ if $z<\hbl$.
%
We distinguish between two cases where $h>\hbl$ (Regime 1) and $h<\hbl$ (Regime 2).  To make analytical progress, we specialize to the step function in Equation~\eqref{eq:transition}.  This is a drastic simplification, but it does simplify the analysis in both regimes considerably.

\subsection{Regime 1}

When we use the step function in Equation~\eqref{eq:transition}, the kinematic condition for $h$ (in case when $h>\hbl$) takes on a simple form:
\begin{equation}
\frac{\partial h}{\partial t}+u_o(r,t)\frac{\partial h}{\partial r}=-\frac{2}{t+t_0}(h-\hbl).
\label{eq:dhdt1}
\end{equation}
By the method of characteristics, this equation has solution:
\begin{equation}
h(r,t)=\frac{1}{(t+t_0)^2}g\left(\frac{r}{t+t_0}\right)+h_{PI}(t),
\label{eq:ht1}
\end{equation}
where $g(\cdot)$ is an arbitrary function and $h_{PI}(t)$ is the particular integral
\begin{equation}
h_{PI}(t)= \tfrac{4}{15}\left[\hbl(t) \left(\frac{3t+5t_0-2t_1}{t+t_0}\right)   -\hbl(\tau) \left(\frac{3\tau+5t_0-2t_1}{\tau+t_0}\right)\right].
\label{eq:ht2}
\end{equation}
Based on these assumptions, we write down a formula linking the total volume of the rim-lamella structure ($V_{tot}$), the rim volume $V$, and the lamella volume $2\pi\int_0^R rh(r,t)\,\mathd r $:
\begin{equation}
V_{tot}=V+2\pi\int_0^R rh(r,t)\mathd r.
\end{equation}
The volume $V_{tot}$ is conserved and equal to $(4/3)\pi R_0^3$.  Hence, by direct differentiation, we get:
\begin{equation}
\frac{\mathd V}{\mathd t}=2\pi R  \left[ u_o\left(h-\hbl\right) - Uh\right],
\label{eq:volume1}
\end{equation}
where $u_o$ and $h$ are evaluated at $r=R$. 
\begin{remark}
Equation~\eqref{eq:volume1} is true for arbitrary height profiles $g(\cdot)$.
\end{remark}
%In fact, for the present purposes, it will be more useful to work with an equation for the volume, divided by angle, $v=V/(2\pi)$.  
%%
%%
%\begin{equation}
%\frac{\mathd v}{\mathd t}= R  \left[ u_o\left(h-\hbl\right) - Uh\right],
%\label{eq:volume2}
%\end{equation}
%In this way, the volume contained in a small radial sector with angle $\Delta\theta$ is $v\Delta\theta$.

To derive a momentum equation, we look at the total momentum in a small radial sector which sweeps out an angle $\Delta\theta$.  We denote this as $p_{tot}\Delta\theta$; this is made up 
by a contribution from the rim $p\Delta\theta$, and a contribution from the lamella, which is obtained by integration as follows:
%In the same way, we use $p_{tot}\Delta\theta$ to denote the total momentum in the same sector, and we have:
%
\begin{equation}
p_{tot}\Delta\theta=p\Delta\theta+\underbrace{\Delta\theta\rho \int_0^R r u_o(r,t)\int_0^{h(r,t)}F(z)\mathd z}_{=p_{lam}\Delta\theta}.
%p_{tot}=p+\underbrace{\rho \int_0^R r u_o(r,t)\int_0^{h(r,t)}F(z)\mathd z}_{=p_{lam}}.
\end{equation}
We apply Equation~\eqref{eq:nval} and the step-function approximation ($n=1$) to this equation.  We further cancel the factor of $\Delta\theta$ to obtain:
\begin{equation}
p_{tot}=p+\underbrace{\rho \int_0^R r\,\mathd r\, u_o(r,t)\left[h(r,t)-\hbl(t)\right]}_{=p_{lam}}.
\end{equation}
Direct differentiation of $p_{lam}$ then gives:
\begin{equation}
\frac{\mathd p_{lam}}{\mathd t}=\rho R\left[\frac{R\dot R}{t+t_0}-\frac{R^2}{(t+t_0)^2}\right]\left(h-\hbl\right)-\tfrac{1}{6}\rho\frac{R^3}{t+t_0}\frac{\hbl}{t+t_1}.
\label{eq:Plam1}
\end{equation}
Here, we have used the overdot notation for time differentiation: $\dot R\equiv \mathd R/\mathd t$.   We also have $U\equiv \dot R$. 

To make further progress, we use Newton's law, $(\mathd p_{tot}/\mathd t)\Delta\theta= f_{ext}\Delta\theta$, where $f_{ext}\Delta\theta$ is the net external force on the sector.  Hence, for the rim, we have $\mathd p/\mathd t=F_{ext}-\mathd p_{lam}/\mathd t$.  Furthermore, $p=\rho UV(\Delta\theta/2\pi)$.  This gives a closed expression for the acceleration of the rim, $\mathd U/\mathd t$:
\begin{equation}
\rho V\frac{\mathd U}{\mathd t}= 2\pi R\rho\bigg\{\left[u_o(R,t)-U\right]^2\left(h-\hbl\right)+ U^2\hbl\bigg\}
+\tfrac{\pi}{3}\rho R^2u_o(R,t)\frac{\hbl}{t+t_1}+2\pi f_{ext}.
\label{eq:Plam2}
\end{equation}
\begin{remark}
Equation~\eqref{eq:Plam2} is true for arbitrary height profiles $g(\cdot)$.
\end{remark}
We identify the different terms in $f_{ext}$ as follows:
\begin{equation}
 f_{ext}\Delta\theta=-\Delta\theta\, R\surften\left(1-\cos\thetadyn\right)+f_{slip}\Delta\theta .
\end{equation}
Here, $\thetadyn$ is the advancing contact angle.  In practice, this is time dependent, however, we take it to be constant to create a simple model.  The term $f_{slip}$ is due to friction, and arises because the fluid in the rim is moving with a velocity $U$ with respect to the lamella (slip).   Using the  Navier slip model, this term can be treated as:
\begin{equation}
f_{slip}\Delta\theta=-\frac{\mu}{\beta_s} (2a) (R\Delta\theta) U,
\end{equation}
where $\beta_s$ is an effective slip length and $2a$ is the footprint of the rim (See Figure~\ref{fig:schematic}).  Here, we have grouped together the terms $R\Delta\theta$ to highlight the role played by the contact area of the rim and the substrate $(2a)(R\Delta\theta)$.  Putting all this together, we have:
\begin{multline}
\rho V\frac{\mathd U}{\mathd t}= R\rho\bigg\{\left[u_o(R,t)-U\right]^2\left(h-\hbl\right)+U^2\hbl\bigg\}
+\tfrac{\pi}{3}\rho R^2u_o(R,t)\frac{\hbl}{t+t_1}\\
- 2\pi R\surften\left(1-\cos\thetadyn\right)-\frac{\mu}{\beta_s}(2a) (2\pi R) U.
\label{eq:LON0}
\end{multline}
We may also gather up terms to write the momentum flux as a perfect square:
%\begin{multline}
\begin{equation}
\rho V\frac{\mathd U}{\mathd t}= R h \rho \left(\overline{u}-U\right)^2+ R \hbl \rho u_o^2\left[\left(1-\frac{\hbl}{h}\right)+\tfrac{\pi}{3}\frac{t+t_0}{t+t_1}\right]
-2\pi R\surften\left(1-\cos\thetadyn\right)-\frac{\mu}{\beta_s}(2\pi R)(2a) U.
\label{eq:LON1}
\end{equation}
%\end{multline}
%
%
Here, $\overline{u}$ refers to the mean velocity,
\begin{equation}
h\overline{u}=u_o(R,t)\left(h-\tfrac{1}{n}\hbl\right).
\end{equation}
We gather up all the relevant equations in one place:
\begin{subequations}
\begin{eqnarray}
\frac{\mathd V}{\mathd t}&=&2\pi R h \left(\overline{u}-U\right),\\
V\frac{\mathd U}{\mathd t}&=&2\pi R h \rho \left(\overline{u}-U\right)^2+2\pi R\hbl \rho u_o^2\left[\left(1-\frac{\hbl}{h}\right)+\tfrac{1}{6}\frac{t+t_0}{t+t_1}\right]\\
&\phantom{=}&\phantom{aaaaa}
-2\pi R\surften\left(1-\cos\thetadyn\right)-\frac{\mu}{\beta_s}(2\pi  R)(2a) U,\nonumber\\
\frac{\mathd R}{\mathd t}&=&U.
\end{eqnarray}%
\label{eq:RLregime1}%
\end{subequations}%

%I need to justify leaving out the last two terms......
%
%%The term in the underbrace represents the flux of momentum entering the rim through the boundary layer.  Following Reference~\cite{gordillo2019theory}, this term is henceforth neglected, on the basis that its contribution to the acceleration in the rim is negligible compared to the other two terms.    
%%Hence, the final rim-lamella model is closer to the one in Reference~\cite{gordillo2019theory}, rather than that 
%%
%%In practice, the term labelled as $\visc$ can be ignored because it represents a small contribution to the momentum flux, proportional to $\hbl$, whereas the term $2\pi rh \left(\overline{u}-U\right)^2$ involves a momentum flux into the rim proportional to the full lamella height $h$.  We comment further on the validity of this assumption in Section~\ref{sec:theory}, below.  A key point of departure in Equation~\eqref{eq:LON} (compared e.g. to Reference~\cite{eggers2010drop}) is the appearance of the momentum flux term as a perfect square.  This has implications for the well-posedness of the system of equations, which we describe in further detail below, in the context of boundary conditions.  We point out finally 
%%
%We emphasize finally that Equations~\eqref{eq:volume1} and~\eqref{eq:LON1} have different behaviours, depending on whether $h<\hbl$ or $h>\hbl$.  We summarize these two distinct regimes in the following sub-sections.

\subsection{Regime 2}

\begin{figure}
	\centering
		\subfigure[]{\includegraphics[width=0.45\textwidth]{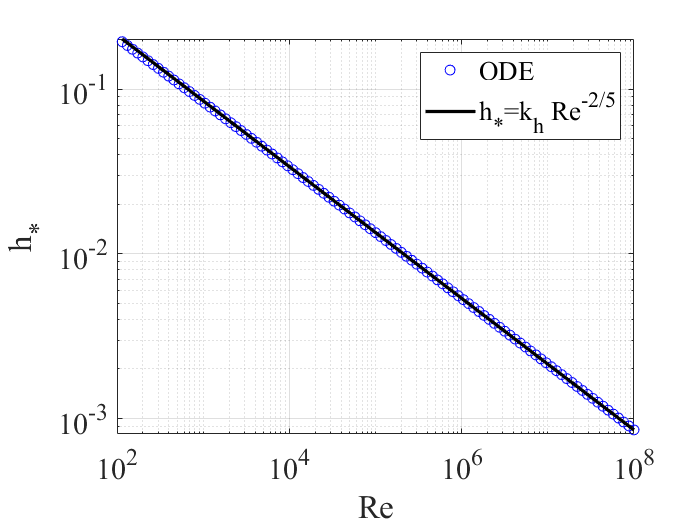}}
		\subfigure[]{\includegraphics[width=0.45\textwidth]{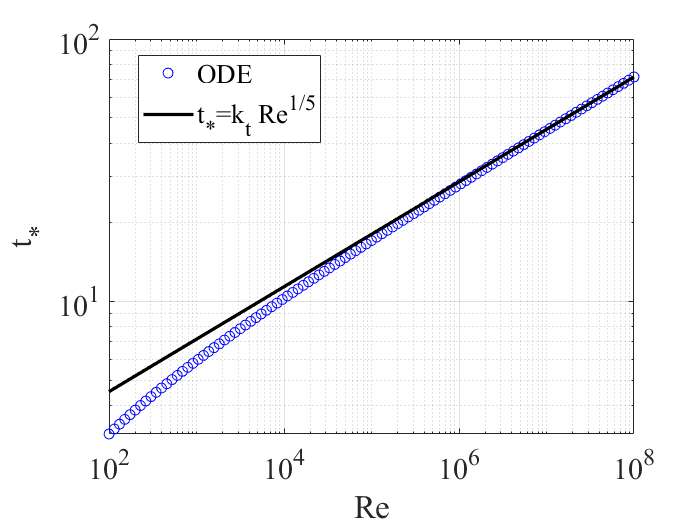}}
		\caption{Plots showing the dependence of $h_*$ and $t_*$ on $\myRe$. Numerical parameter values are given by Equation~\eqref{eq:params}. }
	\label{fig:hconst}
\end{figure}

In the step-function approximation, all flow inside the lamella stops once $h(R,t)$ reaches the boundary-layer height.  We denote this critical time for this by $t_*$, such that $h(R(t_*),t_*)=\hbl(t_*)=h_*$.  Hence, $\overline{u}=0$; by the kinematic condition, $\partial h/\partial t=0$ also.  Thus, the equations of motion~\eqref{eq:LON1} simplify:
\begin{subequations}
\begin{eqnarray}
\frac{\mathd V}{\mathd t}&=&-2\pi R h_* U,\label{eq:star1}\\
V\frac{\mathd U}{\mathd t}&=&2\pi R h_* U^2-2\pi R \surften(1-\cos\thetadyn)-\frac{\mu}{\beta_s}(2\pi  R)(2a) U,\label{eq:star2}\\
\frac{\mathd R}{\mathd t}&=&U.\label{eq:star3}
\end{eqnarray}%
\label{eq:RLregime2}%
\end{subequations}%
To make further progress, we use the `remote asymptotic solution' for $h(R,t)$:
\begin{equation}
h(R,t)=h(\tau)\left(\frac{\tau+t_0}{\tau+t}\right)^2+h_{PI}(t)
\label{eq:ras}
\end{equation}
 which can be obtained from Equation~\eqref{eq:ht1} by assuming that the lamella is flat far from the droplet core, as evidenced by high-speed video analysis of droplet impacts~\cite{roisman2002normal}.  Then, an expression for $t_*$ can be obtained by setting $h(t_*)=\hbl(t_*)=h_*$:
\begin{equation}
\hbl(t_*)=h(\tau)\left(\frac{\tau+t_0}{t_*+t_0}\right)^2+ \tfrac{4}{15}\left[\hbl(t) \left(\frac{3t_*+5t_0-2t_1}{t_*+t_0}\right)   -\hbl(\tau) \left(\frac{3\tau+5t_0-2t_1}{\tau+t_0}\right)\right].
\end{equation}
ince $h_*=a\sqrt{t_*+t_1}$, where $a=\alpha \nu^{1/2}$, we have have  $(h_*/a)^2-t_1=t_*$, hence:
\begin{equation}
h_*=\frac{h(\tau)\left(\tau+t_0\right)^2}{ (h_*/a)^2+(t_0-t_1)}+\tfrac{4}{15}h_* \left[ \frac{3(h_*/a)^2+5(t_0-t_1)}{ (h_*/a)^2+(t_0-t_1)}\right]-\tfrac{4}{15}\hbl(\tau) \left(\frac{3\tau+5t_0-2t_1}{\tau+t_0}\right).
\end{equation}
This is an algebraic (cubic) equation in $h_*$.  For large $\myRe=U_0R_0/\nu$, the cubic equation gives
\begin{equation}
\frac{h_*}{R_0}=\myconst_h \myRe^{-2/5}, \qquad \frac{t_*}{T}=\myconst_t \myRe^{1/5},\qquad \myRe\gg 1.
\label{eq:hconst}
\end{equation}
Here, $\myconst_h$ and $\myconst_t$ are constants, independent of $\myRe$; also, $T=R_0/U_0$ is a timescale.  We elaborate more on the non-dimensionalizaton scheme implicit in this equation below.   The dependencies~\eqref{eq:hconst} are evidenced by Figure~\ref{fig:hconst}.
For reasons that will become apparent in Section~\ref{sec:evaluation}, we use the parameter values
\begin{equation}
\alpha = 1.00,\qquad t_0=0.20,\qquad t_1=0.10.
\label{eq:params}
\end{equation}
This then gives rise to values $k_h=1.35$ and $k_t=1.80$.

Equation~\eqref{eq:RLregime2}  admits a simple analytical solution.  Following standard separation-of-variable techniques, we obtain 
\begin{equation}
\frac{\mathd R}{\mathd t}=U,\qquad U=\pm \sqrt{c_*^2\pm |U_*^2-c_*^2|\left(\frac{V_{tot}-\pi R_*^2h_*}{V_{tot}-\pi R^2h_*}\right)^2},\qquad t>t_*,
\label{eq:Utwo}
\end{equation}
an equation which can be solved by quadratures.
Here, we have used the notation $c_*=\sqrt{\surften(1-\cos\thetadyn)/\rho h_*}$.  We have also used the notation $U_*=U(t_*)$  The behaviour of the model in Regime 2 (including clarification on the choice of signs in Equation~\eqref{eq:Utwo}) is described in more detail below, in Section~\ref{sec:theory}.

\subsection{Regularization}

\begin{figure}
	\centering
		\subfigure[]{\includegraphics[width=0.45\textwidth]{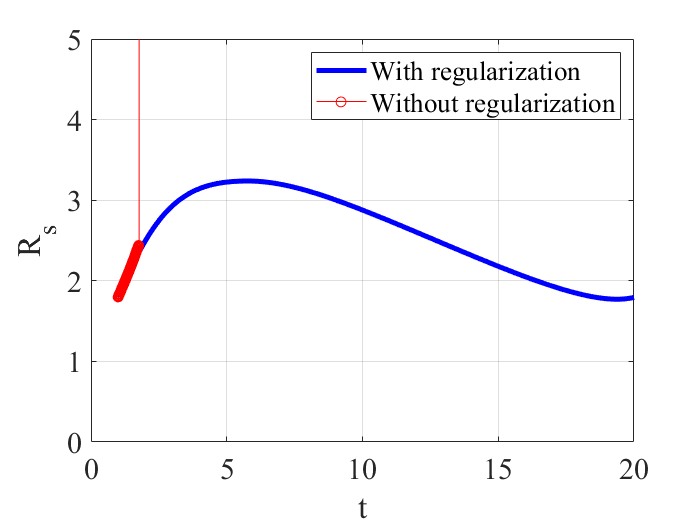}}
		\subfigure[]{\includegraphics[width=0.45\textwidth]{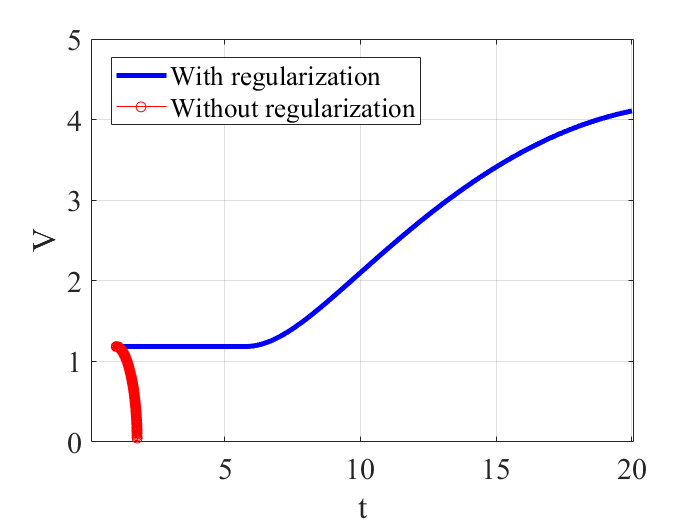}}
		\caption{Effect of the regularization on the rim-lamella model in Reference~\cite{eggers2010drop}.  Panel (a) shows the spreading radius $R_s=R+2a$, where $a$ is computed as  $[V/(2\pi^2R)]^{1/2}$.  Parameter values: $\myRe=\myWe=400$, $\thetadyn=\pi$.  Initial conditions and other parameter values as in the reference.  The initial velocity is chosen to be the same in each case, for a fair comparison between the two cases.}
	\label{fig:eggers_plot_validation0}
\end{figure}

The rim velocity $U$ in Equation~\eqref{eq:RLregime1} can be positive or negative.  If $U$ is positive, the equations of motion admit a singularity, because the volume flux $Q=2\pi R(\overline{u}-U)$ can become negative, leading to $V\rightarrow 0$.  This appears to be implicit in Reference~\cite{eggers2010drop}.  Furthermore, it appears that the authors of that paper have dealt with the issue by  setting the volume flux $Q$ to be zero whenever the rim is advancing (for instance, Figure~10 in that paper has $V=\text{Constant}$ until the rim recedes).  Thus, in this scenario, the singularity in Equation~\eqref{eq:RLregime1} appears to be healed quite simply by setting $V=\text{Constant}$ and hence, $U=[1-({\hbl}/h)]u_o$.  Once $U$ reaches zero (at $t=t_*$), the equations for Regime 2 apply.     A plot  showing the effect of this regularization is given in Figure~\ref{fig:eggers_plot_validation0}.
Without the regularization, the volume touches down to zero, which produces an unphysical infinite acceleration and hence, an infinite radius.
A further plot reproducing some of the results of Reference~\cite{eggers2010drop} is given in Figure~\ref{fig:eggers_plot_validation0}.  This confirms that the regularization implied in that work is indeed in effect.  An advantage of this regularization technique is its simplicity.  Furthermore, it captures the leading-order behaviour $\rr\sim \myRe^{1/5}$.  However, as the Weber number enters the only in the receding phase in this regularized model, the $\myWe$-dependent correction to $\rr$ is not captured in this approach.
%
%xxxxxxxxxxxxxxxxxxxxxxxxxxxxxx
%
%
\begin{figure}
	\centering
		\includegraphics[width=0.6\textwidth]{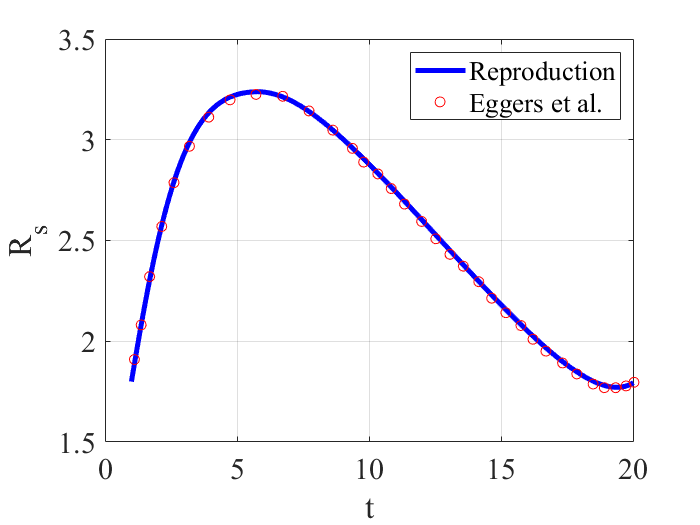}
	\caption{Reproduction of Figure 9 in Reference~\cite{eggers2010drop}, with regularization of the singularity implied in that work.    Parameter values: $\myRe=\myWe=400$, $\thetadyn=\pi$.  Initial conditions and other parameter values as in the reference.  A similar exercise (not shown) has been done to reproduce the reulsts of Figure 11 in Reference~\cite{eggers2010drop}. }
		\label{fig:eggers_plot_validation}
\end{figure}
\begin{figure}
	\centering
		\includegraphics[width=0.6\textwidth]{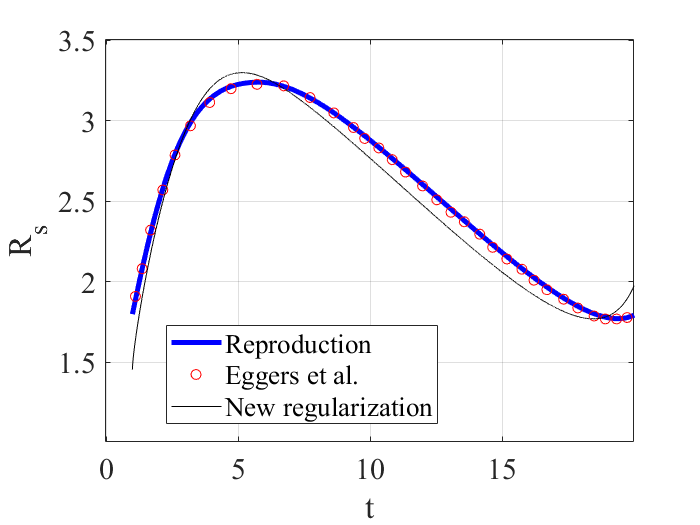}
	\caption{Comparison between the different regularization approaches.  `New Regularization' refers to Equation~\eqref{eq:RLregime1_reg} with initial conditions given below (e.g. Equation~\eqref{eq:sqrt2}).  All other parameter values the same as before. }
		\label{fig:new_reg}
\end{figure}

On the other hand, in Reference~\cite{gordillo2019theory}, the volume-conservation equation $\mathd V/\mathd t=2\pi R(\overline{u}-U)$ is retained throughout the calculations, while a term analogous to $2\pi R\hbl \rho u_o^2[\cdots]$ in Equation~\eqref{eq:RLregime1} is omitted from the momentum equation on the basis that this contribution to the momentum flux is due to the boundary layer and as such, is negligible compared to the contribution from the outer region.  By following this prescription, the singularity in Equation~\eqref{eq:RLregime1} is avoided (e.g. Figure~\ref{fig:new_reg}).  
%Clearly, this term is not negligible, as it is crucial to regularizing the equations of motion.  

A possible physical explanation for the occurrence of this singularity is that both the step-function and piecewise-linear interpolation functions $F(z)$ lead to an underestimation of the momentum in the boundary layer, and therefore, an underestimation of the volume flux $Q$.  The various regularizations then represent an attempt to correct for this underestimation.

A summary of the different models and regularizations is given in Table~\ref{tab:summary}.  The modelling approach taken in the present work is blend of the two previous approaches.  In particular, we omit the boundary-layer contribution to the momentum flux in Regime 1, such that  the final model used herein is given by:
\begin{subequations}
\begin{eqnarray}
\frac{\mathd V}{\mathd t}&=&-2\pi R h \left(\overline{u}-U\right),\\
V\frac{\mathd U}{\mathd t}&=&2\pi R h \rho \left(\overline{u}-U\right)^2-2\pi R\surften\left(1-\cos\thetadyn\right)-\frac{\mu}{\beta_s}(2\pi  R)(2a)U,\\
\frac{\mathd R}{\mathd t}&=&U.
\end{eqnarray}%
\label{eq:RLregime1_reg}%
\end{subequations}%
In Regime 2, the equations are unchanged from before (Equation~\eqref{eq:RLregime2}).  
We henceforth neglect the slip contribution $-(\mu/\beta_s)(2\pi a R) U$ on the basis that the known correlations for the spreading radius (which  Equation~\eqref{eq:RLregime1_reg} must reproduce) depend only on $\mu$ and $\sigma$, through the parameters $\myRe$ and $\sigma$, and as such are independent of slip length, at least in a first approximation.

The approaches to the regularization in Table~\ref{tab:summary} are somewhat ad hoc.  A justification for the present approach is that it gives results that are in agreement with the experimental data. 
However, more work is needed in future to produce a simple rim-lamella model that is intrinsically well-posed, i.e. without having to resort to ad-hoc regularization.
\begin{table}
	\centering
		\begin{tabular}{|p{3cm}|p{3cm}|p{5cm}|p{3.5cm}|}
		\hline
			   & $F(z)$ & Regularization & Friction Force\\
				\hline
				\hline
				Eggers \textit{et al.}~\cite{eggers2010drop} & Step Function & $Q=0$ when $U>0$ & Not considered\\
				\hline
				Gordillo \textit{et al.}~\cite{gordillo2019theory} & Linear & Boundary-layer contribution to momentum flux set to zero& Deemed negligible\\
				\hline
				This work & Step Function & \phantom{aaaaaaaaa}\dittotikz&\phantom{aaaaa}\dittotikz \\
				\hline
		\end{tabular}
		\caption{Summary of the different transition functions and regularizations used in the literature}
		\label{tab:summary}
\end{table}

\subsection{The Initial Conditions}

Initial conditions apply at the onset of Regime 1.  The model has built-in an implied set of initial conditions that give rise to rim generation.
 At time $t=R_0/U_0=\tau$, we take the rim volume to be zero.  This is an approximation to simplify the analysis.  However, early in the process of rim generation, the rim volume is small, as confirmed by experiments~\cite{Riboux2014} and simulations~\cite{naraigh2023analysis}.   
%
%For instance, theory and experiments~\cite{Riboux2014} show that the rim volume at the `ejection time' $t_e\sim \myWe^{-2/3}$ scales as $\myWe^{-4/3}$.  Numerical simulations~\cite{naraigh2023analysis} also show that early on in the rim-lamella evolution, the majority of the mass of the rim-lamella complex is in the lamella.  
%
%
Therefore, $V(\tau)=0$ is a crude approximation, albeit a justifiable one.   Furthermore, this implication gives a natural way to `parametrize' the rim generation phenomenon, since, by taking $V(\tau)=0$, we obtain:
\begin{equation}
\left(\overline{u}-U\right)^2h - \frac{\surften\left(1-\cos\vartheta_{ap}\right)}{\rho} =0.
\label{eq:Utau1}
\end{equation}
This equation follows from the rim-lamella model with the condition that the acceleration should be finite at $\tau=0$.  Hence:
\begin{equation}
U(t=\tau)=\overline{u}-\sqrt{ \frac{\surften\left(1-\cos\vartheta_{ap}\right)}{\rho h}}, 
\label{eq:Utau2}
\end{equation}
or
\begin{equation}
U(t=\tau)=\frac{R_{init}}{\tau+t_0}\left(1-\frac{\hbl}{h_{init}}\right)-c(\tau).%\sqrt{ \frac{\surften\left(1-\cos\vartheta_{ap}\right)}{\rho h_{init}}}.
\label{eq:sqrt2}
\end{equation}
%For these reasons, we use Equation~\eqref{eq:Utau1} as the initial condition.  All  quantities on the right  are evaluated at the initial time.  Thus, the initial velocity can be written in more detail as follows:
%%
%\begin{equation}
%U(t=\tau)=\frac{R_{init}}{\tau+t_0}\left(1-\frac{\hbl}{h_{init}}\right)-c(\tau).%\sqrt{ \frac{\surften\left(1-\cos\vartheta_{ap}\right)}{\rho h_{init}}}.
%\label{eq:sqrt2}
%\end{equation}
%%It should be noted that these arguments are robust to the neglect of the term $\visc$ in Equation~\eqref{eq:LON}.  For, $\visc$ has positive sign; as such, it can be combined with $[c(\tau)]^2$ to give an `effective' characteristic speed.  The arguments regarding the sign under the various square roots in Equation~\eqref{eq:sqrt1} and~\eqref{eq:sqrt2} are then effectively the same.

The values $R_{init}$ and $h_{init}$ describe the `initial lamella', i.e. just prior to the formation of the rim.  For the inviscid case, it is important to describe carefully the dependence of these parameters on $\myWe$~\cite{amirfazli2023bounds}.  For the viscous case, this seems less important.  For instance, Roisman et al. determine $R_{init}$ and $h_{init}$ using an energy-budget analysis~\cite{roisman2002normal}, whereas Eggers \textit{et al.} use a geometric argument~\cite{eggers2010drop}.  In these studies, the final dependence of the spreading radius on $\myRe$ and $\myWe$ is insensitive to the approach used to set the initial conditions.  We use a geometric argument here: we assume that the droplet assumes a `pancake' shape at time $\tau$, with initial height $h_{init}=R_0/2$.  By conservation of mass, the maximum extent of this `pancake' shape is $R_0\sqrt{8/3}$.  The point of departure with respect to the work of Eggers \textit{et al.} is that we assume that all mass is assumed to be in the lamella at this time, hence $V=0$, the expression~\eqref{eq:Utau2} for the initial velocity, and   $R_{init}=R_0\sqrt{8/3}$.  

We emphasize finally that the regularization approach pursued in this work (in particular, Equation~\eqref{eq:RLregime1_reg}) is attractive, because it enables us to develop the initial conditions~\eqref{eq:sqrt2}. An advantage of this set of initial conditions is that they revert to the inviscid ones put forward by  Roisman for the inviscid case~\cite{roisman2002normal}, when $\myRe=\infty$.

%These are an obvious generalization of the initial conditions put forward by  Roisman for the inviscid case~\cite{roisman2002normal}.  In contrast, the initial conditions in the regularization of   

\subsection{Non-dimensionalization}

We henceforth non-dimensionalize the mathematical model~\eqref{eq:RLregime1} and~\eqref{eq:RLregime2} based on the initial droplet radius $R_0$, the pre-impact droplet velocity $U_0$, and the liquid density $\rho$.  This gives a standard timescale $T=R_0/U_0$.  The key dimensionless variables are therefore $\widetilde{R}=R/R_0$, $\widetilde{h}=h/R_0$, and $\widetilde{U}=U/U_0$.  Equations~\eqref{eq:RLregime1} and~\eqref{eq:RLregime2} are then replaced with dimensionless analogues: with the formal replacements $\rho\rightarrow 1$, $\nu\rightarrow 1/\myRe$, and $\surften\rightarrow 1/\myWe$, where $\myRe$ and $\myWe$ are the Weber number and Reynolds number, given previously in Equation~\eqref{eq:maindefs}.
Correspondingly, the boundary-layer thickness becomes:
\begin{equation}
\widetilde{\hbl}=\frac{\hbl}{R_0}=\alpha \myRe^{-1/2}\sqrt{\widetilde{t}+\widetilde{t_1}},
\end{equation}
where $\widetilde{t}=t/T$ is the dimensionless time variable and $\widetilde{t_1}=t_1/T$ is a constant.  Similarly, the
characteristic speed becomes:
\begin{equation}
\widetilde{c}=\frac{c}{U_0}=\sqrt{\frac{1}{\myWe}\frac{1-\cos\thetadyn}{\widetilde{h}}}.
\end{equation}
Following standard practice, we henceforth omit the tildes over the dimensionless variables, it being understood that we work entirely with the dimensionless version of the mathematical model.

\section{Theoretical Analysis}
\label{sec:theory}

In this section we characterize the solutions of the rim-lamella model in both Regime 1 and Regime 2, using some simple concepts from applied analysis and dynamical systems theory.  The aim here is to develop some preliminary results which can be used later on to establish the bounds~\eqref{eq:bounds}.

\subsection{Regime 1}

We analyse the rim-lamella model in Regime 1, given by Equations~\eqref{eq:RLregime1}.  Following previous work on the inviscid case~\cite{amirfazli2023bounds} we introduce the velocity defect $\Delta$.  In the present context, the velocity defect must account for the viscous boundary layer.  Hence, we define
\begin{equation}
\Delta=\overline{u}-U=\frac{R}{t+t_0}\left(1-\frac{\hbl}{h}\right)-U.
\end{equation}
We substitute this expression for $\Delta$ into Equation~\eqref{eq:RLregime1}.  After careful calculations, we obtain an Ordinary Differential Equation for $\Delta$:
\begin{equation}
\frac{\mathd \Delta}{\mathd t}+\frac{\Delta}{t+t_0}\left(1-\frac{\hbl}{h}\right)=-\frac{2\pi R h}{V}\left(\Delta^2-c^2\right)-\frac{R}{(t+t_0)^2}\frac{\hbl}{h}\underbrace{\left[3\left(1-\frac{\hbl}{h}\right)+\tfrac{1}{2}\frac{t+t_0}{t+t_1}\right]}_{=\Phi(t)}.
\label{eq:dDeltadt}
\end{equation}
Here, $\Phi(t)\geq 0$ for $t\leq t_*$.  We further introduce $Y=\Delta-c$.  By direct calculation, we obtain:
\begin{equation}
\frac{\mathd Y}{\mathd t}+\frac{Y}{t+t_0}\left(1-\frac{\hbl}{h}\right)+\frac{4\pi R h c}{V}Y=\underbrace{-\frac{2\pi R h}{V}Y^2-\frac{2c}{t+t_0}\left(1-\frac{\hbl}{h}\right)-\frac{R}{(t+t_0)^2}\frac{\hbl}{h}\Phi(t)}_{\leq 0}.
\label{eq:dYdt}
\end{equation}
We identify the integrating factor
\begin{equation}
\mu=\exp\left[\int\frac{1}{t+t_0}\left(\frac{h-\hbl}{h}\right)\mathd t+\int \frac{4\pi R h c}{V}\mathd t\right].
\end{equation}
We use the remote asymptotic solution~\eqref{eq:ras} to write $\mathd h/\mathd t=-2(t+t_0)^{-1}(h-\hbl)$ and hence,
\begin{equation}
\mu= h^{-1/2}\,\mathe^{\int (4\pi R hc/V)\mathd t}.
\end{equation}
By Gronwall's Inequality, Equation~\eqref{eq:dYdt} yields:
\begin{equation}
\frac{Y(t)}{[h(t)]^{1/2}}\leq \frac{Y(\tau)}{[h(\tau)]^{1/2}}\mathe^{-\int_t^\tau (4\pi R hc/V)\mathd t}
\end{equation}
Since the exponential term is less than or equal to one, we have:
\begin{equation}
Y(t)\leq \left[ \frac{h(t)}{h(\tau)}\right]^{1/2}Y(\tau)=\left[ \frac{h(t)}{h(\tau)}\right]^{1/2}\left[\Delta(\tau)-c(\tau)\right].
\end{equation}
In view of the given initial conditions, this gives $Y(t)\leq 0$, hence:
\begin{equation}
\Delta(t)\leq c(t),\qquad t\in [\tau,t_*].
\label{eq:Deltaineq1}
\end{equation}

 Another way to arrive at the result~\eqref{eq:Deltaineq1} is to examine the behaviour of the original Equation~\eqref{eq:dDeltadt} in the neighbourhood of $c(t)$.  Hence, we write $\Delta=c(t)-\epsilon(t)$, where $\epsilon(t)$ is small and positive.  We substitute this approximation into Equation~\eqref{eq:dDeltadt} and neglect quadratic terms in $\epsilon$.  This gives:
\begin{equation}
\frac{\mathd \epsilon}{\mathd t}+\frac{\epsilon}{t+t_0}\left(1-\frac{\hbl}{h}\right)=-\epsilon \frac{4\pi Rh c}{V}+\underbrace{\left[\frac{2c}{t+t_0}\left(1-\frac{\hbl}{h}\right)+\frac{R}{(t+t_0)^2}\frac{\hbl}{h}\Phi(t)\right]}_{\geq 0}+O(\epsilon^2),
\end{equation}
with $\epsilon(t)\geq 0$.  Using Gronwall's Inequality again, one can show that if $\epsilon(\tau)=0$ initially, then $\epsilon(t)\geq 0$ thereafter, and hence, $\Delta(t)\leq c(t)$, $ t\in [\tau,t_*]$.

The foregoing results follow the same pattern established previously in the inviscid case~\cite{amirfazli2023bounds}.  There, it was also possible to establish that $\Delta(t)>-c(t)$.  However, it is not possible to derive an analogous result for the viscous case.  For, by writing $\Delta(t)=-c(t)+\epsilon(t)$ in Equation~\eqref{eq:dDeltadt} and assuming that $\epsilon(t)$ is small, one obtains:
\begin{equation}
\frac{\mathd \epsilon}{\mathd t}+\frac{\epsilon}{t+t_0}\left(1-\frac{\hbl}{h}\right)=\epsilon \frac{4\pi Rh c}{V}+\underbrace{\left[\frac{2c}{t+t_0}\left(1-\frac{\hbl}{h}\right)-\frac{R}{(t+t_0)^2}\frac{\hbl}{h}\Phi(t)\right]}+O(\epsilon^2).
\end{equation}
The term in the underbrace is not sign-definite, and hence, it is possible for $\epsilon$ to go below zero.  When this happens, we get $V\mathd U/\mathd t=0$, and a finite-time singularity with $V\rightarrow 0$ cannot be ruled out.  By analyzing the prefactors in the problematic term (e.g. $c(t)\sim \myWe^{-1/2}$, $(\hbl/h)\Phi(t)\sim \myRe^{-1/2}$), it can be seen that this same term is likely to become negative when $\myRe<\myWe$, or $\mathrm{Ca}=\mu U_0/\sigma>1$.  In this regime, a potential cure for the singularity is to restore the slip term to Equation~\eqref{eq:RLregime1_reg}.

%\begin{remark}
%The foregoing results are not affected by the inclusion of the term $\visc$ in the equation for $\mathd U/\mathd t$.  For, including this term would  amount to an additional term in Equation~\eqref{eq:dDeltadt}, with the same sign as $\Phi(t)$.  In the foregoing analysis, the term involving $\Phi(t)$ is dealt with using inequalities, and the inequalities are not affected by adding in $\visc$.
%\end{remark}

\subsection{Regime 2}

\begin{figure}
	\centering
		\includegraphics[width=0.7\textwidth]{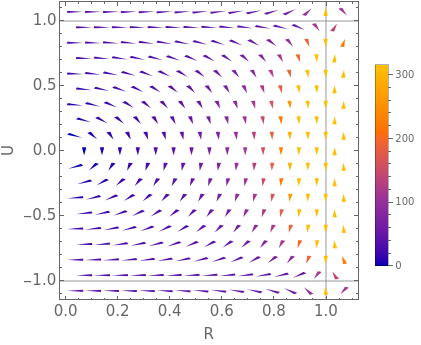}
		\caption{ Phase portrait of the dynamical system~\eqref{eq:dynsys}. The parameters $c_*$ and $V_{tot}/\pi h_*$ have been taken to be unity for illustration purposes. The color code indicates the strength of the flow vector field of the dynamical system. The lines $R=1$ and $U=\pm 1$ indicate the separatrices of the dynamical system.  The line $R=1$ corresponds to the rim volume going to zero. }
	\label{fig:phase}
\end{figure}

We also investigate the rim-lamella model~\eqref{eq:RLregime2} in Regime 2.  Using $V=V_{tot}-\pi R^2h_*$ in Regime 2, the equations can be reduced to a pair of equations in the variables $(R,U)$:
\begin{equation}
\left(V_{tot}-\pi R^2 h_*\right)\frac{\mathd U}{\mathd t}=2\pi Rh_*\left( U^2-c_*^2\right),\qquad \frac{\mathd R}{\mathd t}=U.
\label{eq:dynsys}
\end{equation}
This can be further identified as a two-dimensional dynamical system, $(\mathd /\mathd t)(R,U)=\bm{F}(R,U)$, where $\bm{F}$ is a two-dimensional vector field.  A phase portrait showing the flow $\bm{F}$ is provided in Figure~\ref{fig:phase}.
From the figure, it can be seen that all trajectories starting at $U_*<c_*$  either tend to a closed periodic orbit, or are attracted to $U\rightarrow -c_*$, as $t\rightarrow \infty$.  Trajectories starting at $U_*>c_*$ diverge.  This divergence corresponds to the  singularity described in Section~\ref{sec:theory}.  Referring back to that section, the divergence is avoided if $\Delta(t)>-c(t)$ for all $t$ in Regime 1.  

We emphasize that the trajectories in Figure~\ref{fig:phase} with $U_*<c_*$ flow into negative values of $R$ at late times.  This is of course unphysical, and signals the end of the validity of the rim-lamella model.  However, those segments of a trajectory with $R>0$ are physical.  In particular,  the `first-quarter' of a closed trajectory is physically sensible, and may give insights into the timescale associated with droplet retraction.  For this reason, we compute an estimate for the quarter-period $T_4$ of such a closed trajectory in what follows.

%We start from ``regime 2'' equations, assuming we are in the region $\Omega \subset \mathbb{R}^2$ of the $(R,U)$ plane defined by the inequalities $U^2-c_*^2 < 0$ and $0 < V_{tot} -\pi R^2 h_*$. This assumption will be justified once we obtain the Hamiltonian. The constants $V_{tot}, h_*$ and $c_*$ are all positive.
%\begin{equation}
% \frac{\mathrm{d}R}{\mathrm{d}t} = U\,, \qquad \frac{\mathrm{d}U}{\mathrm{d}t} = \frac{2\pi R h_*(U^2-c_*^2)}{V_{tot} -\pi R^2 h_*}\,.
%\end{equation}

\subsection{Regime 2: Analytical Computation of the Quarter-Period}

We start by reducing the dynamical system~\eqref{eq:dynsys} to a Hamiltonian form.  To do this, we seek a Jacobi last multiplier $M(R,U)$ such that:
%The above dynamical system is autonomous and $2$-dimensional, so if one can find a Jacobi last multiplier then the system is Hamiltonian with canonical Poisson bracket. Such a multiplier $\rho(R,U)$ is found by requesting that the divergence of the flow field be zero:
\begin{equation}
\frac{\partial}{\partial R}\left( M U \right) +  \frac{\partial}{\partial U}\left(M  \frac{2\pi R h_*(U^2-c_*^2)}{V_{tot} -\pi R^2 h_*}\right) = 0.
\end{equation}
By inspection we find that $M= {(c_*^2-U^2)}^{-1}$ solves this equation, so we obtain the Hamiltonian system
\begin{equation}
 \rho \frac{\mathrm{d}R}{\mathrm{d}t} = \frac{U}{c_*^2-U^2} =: \frac{\partial H}{\partial U}\,, \qquad \qquad \rho \frac{\mathrm{d}U}{\mathrm{d}t} = -\frac{2\pi R h_*}{V_{tot} -\pi R^2 h_*} =: -  \frac{\partial H}{\partial R}\,,
\end{equation}
and integrating these equations we get the Hamiltonian
\begin{equation}
H(R,U) = -\frac{1}{2}\log (c_*^2-U^2)-\log(V_{tot} -\pi R^2 h_*).
\end{equation}
By inspection, the domain $\Omega$ defined by the inequalities $c_*^2-U^2>0$ and $0<V_{tot} -\pi R^2 h_*$ is invariant under the evolution, as the Hamiltonian is singular at the boundary $\partial \Omega$.  Thus, this analysis remains valid provided $c_*^2-U^2>0$ and $0<V_{tot} -\pi R^2 h_*$ at the onset of Regime 2.   The first inequality is satisfied for trajectories that start at $0<U_*<c_*$.  The second inequality is satisfied automatically, provided the rim volume is non-zero at the onset of Regime 2, since $V_{tot}=V+\pi R^2h_*$.

In terms of the positive constant $\beta := \exp(-H)$ we get 
\begin{equation}
\beta^2 = (c_*^2-U^2)(V_{tot} -\pi R^2 h_*)^2 \quad \Longrightarrow \quad  U^2=c_*^2-\frac{\beta^2}{(V_{tot} -\pi R^2 h_*)^2},
\end{equation}
and recalling that $U=\mathd R/\mathd t$ we get
\begin{equation}
\left(\frac{\mathrm{d}R}{\mathrm{d}t}\right)^2=c_*^2-\frac{\beta^2}{(V_{tot} -\pi R^2 h_*)^2}.
\end{equation}
Thus, we have reduced the problem to an effective 1-dimensional particle with a potential given by a rational polynomial. Before proceeding any further with the solution of this, it is useful to change to non-dimensional variables via the transformations
\begin{equation}
\label{eq:adim_transf}
\newt:= t c_* \sqrt{\frac{\pi h_*}{V_{tot}}} \qquad \newR := R \sqrt{\frac{\pi h_*}{V_{tot}}}\,,
\end{equation}
which transforms the equations to the equivalent formulation
\begin{equation}
\label{eq:adim_evol}
\left(\frac{\mathd\newR}{\mathd \newt}\right)^2=1-\frac{\newconst^2}{(1 -\newR^2)^2} = \frac{(1 -\newR^2)^2-\newconst^2}{(1 -\newR^2)^2}\,,
\end{equation}
where $\newconst$
%=B/(c_*V_{tot})$ 
is another constant. The original domain $\Omega$ and the non-negativity of $(\mathd \newR/\mathd \newt)^2$ simply map to  the condition $0<\newconst<1-\newR^2$ for the constant $\newconst$ and the variable $\newR$.
Therefore, the evolution is reduced to the above system \eqref{eq:adim_evol} with the free parameter $0<\newconst<1$, and the variable $\newR$ is restricted by the condition $\newR^2 < 1-\newconst$.
The dynamics of the variable $\newR$ is very simple: given $\newconst$, the motion is periodic. Due to the symmetries of  equation \eqref{eq:adim_evol} under $\newR\to -\newR$ and under $\newt \to -\newt$, the relevant time scale is the quarter period $\newT_4$, corresponding to the evolution in the first quadrant $\newR\geq 0$, $\mathd \newR/\mathd\newt \geq 0$.
We get, taking square roots of both sides of \eqref{eq:adim_evol} and integrating the differential $\mathd\newt$,
\begin{equation}
\newT_4(\newconst) = \int_0^{\sqrt{1-\newconst}} \frac{1-\newR^2}{\sqrt{(1-\newR^2)^2-\newconst^2}}\mathrm{d}\newR = \frac{(\newconst+1)E\left(\frac{1-\newconst}{\newconst+1}\right)-DK\left(\frac{1-\newconst}{\newconst+1}\right)}{\sqrt{\newconst+1}},
\end{equation}
where $E(m)$ is the complete elliptic integral and $K(m)$ is the complete elliptic integral of the first kind.
It is easy to check that the dependence of $\newT_4(\newconst)$ on the parameter $\newconst$ is very slight in the relevant domain $0<\newconst<1$, varying monotonously from $\newT_4(0)=1$ to $\newT_4(1) = \pi/2\sqrt{2}\approx 1.111$.
Thus, going back to the transformations~\eqref{eq:adim_transf}, the quarter period in physical units is
\begin{equation}
T_4 =  \sqrt{\frac{V_{tot}}{\pi h_* c_*^2}} \,\newT_4(\newconst) \approx \sqrt{\frac{V_{tot}}{\pi h_* c_*^2}},
\end{equation}
Correspondingly, the maximum value of $R$ achievable in Regime 2 is $R_{max}^{(2)}$, which is attained when $\newR = \sqrt{1-\newconst}$.  Thus,
\begin{equation}
R_{\max}^{(2)} = \sqrt{\frac{V_{tot}}{\pi h_*}} \sqrt{1-\newconst} \leq \sqrt{\frac{V_{tot}}{\pi h_*}}.
\label{eq:Rmaxtwo}
\end{equation}
%
%which is maximised in the limit $\newconst\to 0$. 
%
Finally, in terms of the dependence of these constants on the Reynolds and Weber numbers, we get, from $h_* \sim \myRe^{-2/5}$ and $h_*c_*^2 \sim \myWe^{-1}$,
\begin{equation}
T_4 \sim \myWe^{1/2}\,, \qquad R_{\max}^{(2)} \sim \myRe^{1/5}.
\end{equation}
where we have assumed that $V_{tot}$ has no dependence on these numbers.

These results indicate that rim-lamella model specialized to Regime 2 gives a good rough approximation to the dynamics.  In particular, $R_{\max}^{(2)} \sim \myRe^{1/5}$ reproduces the first term in the empirical correlation~\eqref{eq:correlation}.  Also, $T_4\sim \myWe^{1/2}$ indicates that the time for significant retraction to occur is dominated by surface-tension effects.

% Thus, these can constituted using whatever kind of pancakeology you like, e.g. with $h_{init}=R_0/2$, we get $R_{init}=\sqrt{8/3}R_0$ ($R_0$ is the initial droplet radius, prior to impact).

\section{Bounds on the spreading radius}
\label{sec:bounds}

In this section we derive explicit bounds for $\rmax=\max_{t\geq \tau}R(t)$.  The aim is to show that there exist bounds
\begin{equation}
f(\myRe,\myWe)\leq \rmax \leq g(\myRe,\myWe),
\label{eq:bounds1}
\end{equation}
valid for appropriate ranges of $\myRe$ and $\myWe$.  

%The strategy is to introduce further quantities,
%%
%%
%%
%\begin{equation}
%\rmax=\max_{\stackrel{\text{Regime 1}}{t\in[\tau,t_*]}}R(t),\qquad R_*=R(t_*).
%\end{equation}
%to establish that $\rmax \geq R_*$ (for suitable $\myRe$ and $\myWe$) and from there, to infer the bounds~\eqref{eq:bounds}.  Figure~\ref{fig:} shows a graphical description of the strategy.
%

\subsection{Upper bound}

We use a simple geometric argument to derive an upper bound on $\rmax$.  We have:
\begin{equation}
\pi \rmax^2 h(t_{max})=V_{tot}-V(t_{max})\leq V_{tot}.
\end{equation}
We use the principle of mass conservation and the initial conditions in Section~\ref{sec:model} to obtain $V_{tot}=\pi R_{init}^2 h_{init}$.  Hence:
\begin{equation}
\rmax \leq R_{init}h_{init}^{1/2} [h(t_{max})]^{-1/2}.
\label{eq:Rub}
\end{equation}
There are now two cases to consider.

\paragraph*{Case 1.}  The maximum occurs for $t_{max}\leq t_*$.  Then, since $h$ is a monotone-decreasing function, we have $h(t_{max})\geq h(t_*)$, hence
$[h(t_{max})]^{-1/2}\leq [h(t_*)]^{-1/2}$, hence
\begin{equation}
\rmax \leq R_{init}h_{init}^{1/2} [h(t_*)]^{-1/2}.
\label{eq:Rub1}
\end{equation}
\paragraph*{Case 2.}  The maximum occurs for $t_{max}\geq t_*$. Since $h(t)=h_*$ for all $t\geq t_*$, the same reasoning as before applies, and the bound~\eqref{eq:Rub} is retained.  Hence, the bound~\eqref{eq:Rub} is valid in both cases.

\begin{remark}
The bound in Regime 2 can also be obtained by consideration of the analytical solution for that regime, in particular, Equation~\eqref{eq:Rmaxtwo}.  
%By setting $U=0$ at $t=t_{max}$ in that equation, we obtain:  
%\begin{equation}
%\pi \rmax^2 h_*=V_{tot}-\frac{\left|U_*^2-c_*^2\right|}{c_*^2}V_*\leq V_{tot}.
%\end{equation}
%With $V_{tot}=\pi R_{init}^2h_{init}$, the bound~\eqref{eq:Rub} is recovered.  
\end{remark}

\subsection{Lower bound}

We develop here an expression $R_*\geq \text{(Lower Bound)}$.  This immediately establishes a lower bound on $\rmax$, since
\begin{equation}
\rmax \geq R_* \geq \text{(Lower Bound)}.
\end{equation}
For this purpose, we refer back to the velocity defect $\Delta$:
\begin{equation}
\frac{\mathd R}{\mathd t}-\frac{R}{(t+t_0)}\left(1-\frac{\hbl}{h}\right)=-\Delta \stackrel{\text{\eqref{eq:Deltaineq1}}}{\geq} -c(t). 
\label{eq:Rineq1}
\end{equation}
This can be re-written as:
\begin{equation}
\frac{1}{h^{1/2}}\frac{\mathd }{\mathd t}\left(R h^{1/2}\right)\geq  -c(t),
\label{eq:Rineq2}
\end{equation}
or
\begin{equation}
\frac{\mathd }{\mathd t}\left(R h^{1/2}\right)\geq  -c(t)h^{1/2}=-\sqrt{\frac{1}{\myWe}(1-\cos\thetadyn)}.
\label{eq:Rineq2}
\end{equation}
By Gronwall's Inequality, we have:
\begin{equation}
R(t_*)\geq R_{init}h_{init}^{1/2} [h(t_*)]^{-1/2}-\sqrt{\frac{1}{\myWe}(1-\cos\thetadyn)}[h(t_*)]^{-1/2}(t_*-\tau).
\end{equation}
Summarizing, our main result so far is the following:
\begin{theorem}
Suppose that the initial conditions in Section~\ref{sec:model} hold.  Then,
\begin{equation}
R_{init}h_{init}^{1/2} [h(t_*)]^{-1/2}-\sqrt{\frac{1}{\myWe}(1-\cos\thetadyn)}[h(t_*)]^{-1/2}(t_*-\tau)\leq 
\rmax
\leq R_{init}h_{init}^{1/2} [h(t_*)]^{-1/2}.
\label{eq:golden1}
\end{equation}
\end{theorem}
Equation~\eqref{eq:golden1} reflects the idea that the maximum spreading radius suffers a correction due to the retraction of the rim.  This is the main idea behind~\eqref{eq:correlation}; our result~\eqref{eq:golden1} is another take on this observation, and is derived here \textit{a priori}, based on the equations of motion~\eqref{eq:RLregime1} only.

\section{Evaluation of the Bounds}
\label{sec:evaluation}

In this section we evaluate the bounds~\eqref{eq:golden1}.  Specifically, we wish to see how sharp the bounds are, in the sense of how well the bounds agree with the numerical solutions of the rim-lamella model.  A second aim of this section is to compare the prediction of the maximum spreading radius in the rim-lamella model, and the energy-budget method, with a view to shedding light on the head-loss phenomenon in the latter.  We will also compare our predictions with the correlation~\eqref{eq:correlation}, which was obtained from experiments.

For these purposes, it will be useful to convert the bounds~\eqref{eq:golden1} into expressions that involve $\myWe$ and $\myRe$ explicitly.  We use
$h(t_*)=\myconst_h \myRe^{-2/5}$ and $t_*=\myconst_t \myRe^{1/5}$ at sufficiently large $\myRe$ (Equation~\eqref{eq:hconst}) to re-write the bounds~\eqref{eq:golden1} as:
\begin{equation}
R_{init}h_{init}^{1/2} \myconst_h^{-1/2}\myRe^{1/5}-\sqrt{(1-\cos\thetadyn)}\myconst_h^{-1/2}\myconst_t \myWe^{-1/2}\myRe^{2/5}\leq 
\rmax
\leq R_{init}h_{init}^{1/2}\myconst_h^{-1/2}\myRe^{1/5},\qquad \myRe\gg 1.
\end{equation}
We gather up constants in an obvious way to write this as:
\begin{equation}
\myconst_1 \myRe^{1/5}-\sqrt{(1-\cos\thetadyn)}\myconst_2 \myWe^{-1/2}\myRe^{2/5}\leq 
\rmax
\leq \myconst_1 \myRe^{1/5},\qquad \myRe\gg 1.
\label{eq:golden2}
\end{equation}
Obviously, for the lower bound to yield any meaningful information, we require $\myWe^{1/2}\gg  \myRe^{1/5}$.  

The results so far have involved expressions for $R_{max}$, the maximum extent of the lamella.  However, what is of interest is $\rr$, being the maximum spreading radius of the rim-lamella structure.  The maximum spreading radius can be written as $\rr=R_{max}+2a$, where $2a$ is the footprint of the rim, and where $a$ can be estimated as~\cite{amirfazli2023bounds}:
\begin{equation}
a=\sin\thetadyn\sqrt{\frac{V}{\pi R \left[2\thetadyn-\sin(2\thetadyn)\right]}}.
\end{equation}
Since $V\leq V_{tot}$ and 
$R_{max}^{-1}\leq [\myconst_1 \myRe^{1/5}-\sqrt{(1-\cos\thetadyn)}\myconst_2 \myWe^{-1/2}\myRe^{2/5}]^{-1}$, the correction $a$ does not exceed a term proportional to $\myRe^{-1/10}$ and hence, can be ignored in the remaining calculations.  For the avoidance of doubt, in the remainder of this section we use $\rr$ as the quantity of interest, as this is what is investigated in the relevant references.  

Throughout this section, we use the numerical parameter values for $\alpha$, $t_0$ and $t_1$ in Equation~\eqref{eq:params}.  This gives $k_1=R_{init}h_{init}^{1/2}/k_h^{1/2}=1.00$.  Thus, an element of fitting is involved in this choice, as this gives $\rr/R_0 \sim 1.00 \myRe^{1/5}$, in (deliberate) agreement with Equation~\eqref{eq:correlation}.  Thus, by choice of these parameters, agreement between the rim-lamella model and the experiments is maximized.

\begin{figure}
	\centering
		\includegraphics[width=0.7\textwidth]{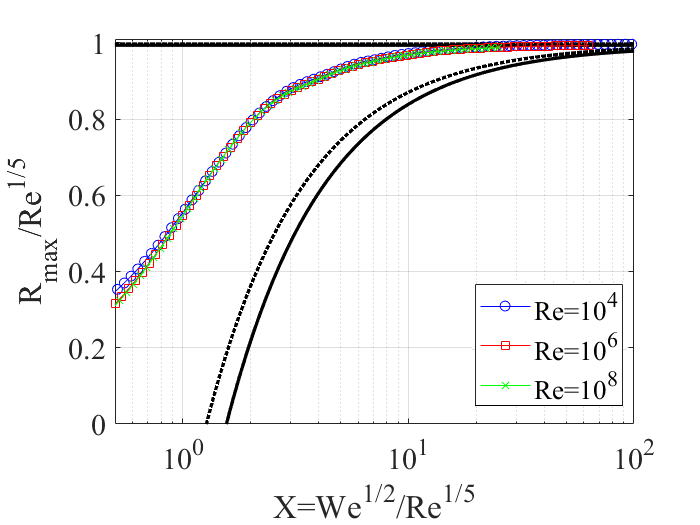}
		\caption{Evaluation of the upper and lower bounds for various values of $\myRe$. Broken line: Bounds~\eqref{eq:golden1}.  Solid line: Bounds~\eqref{eq:golden2}, which admit the explicit dependence on $\myRe$ and $\myWe$.    Advancing contact angle: $\thetadyn=\pi/2$.}
	\label{fig:bound_Re}
\end{figure}

To explore the agreement between the semi-empirical correlation~\eqref{eq:correlation} and the rim-lamella model in more detail, we carry out numerical solutions of the latter.  The numerical solutions are generated by solving the rim-lamella model using \texttt{ode89} in Matlab (specifically, we solve Equation~\eqref{eq:RLregime1} and~\eqref{eq:RLregime2}).  A first set of results is shown in Figure~\ref{fig:bound_Re}.  
The figure confirms that the bounds~\eqref{eq:golden1} successfully sandwich the maximum spreading $\rr$ between extreme values.

\begin{figure}
	\centering
		\includegraphics[width=0.7\textwidth]{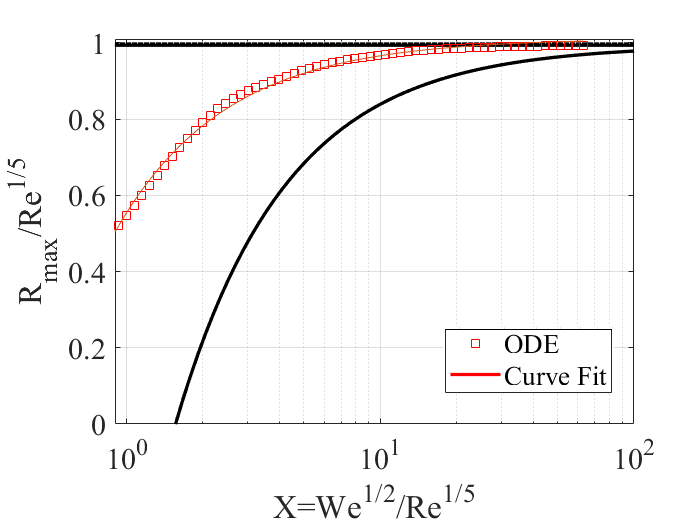}
		\caption{Plot showing $Y=R_{max}/\myRe^{1/5}$ as a function of $X$ using numerical solutions of the rim-lamella model (labelled `ODE' in the legend), the fit $Y=a+bX^{-1}$ to the data, and comparison with the bounds~\eqref{eq:golden2}.    Reynolds number: $\myRe=10^6$.  Advancing contact angle: $\thetadyn=\pi/2$.}
	\label{fig:curve_fitting}
\end{figure}

We also analyze the results in Figure~\ref{fig:bound_Re} in a more quantitative manner.  We introduce the scaled variables
 $X=\myWe^{1/2}/\myRe^{1/5}$ and $Y=\rr/\myRe^{1/5}$.  Hence, we fit the model curve $Y=a+b X^{-1}$ to the data emanating from the numerical solutions of the rim-lamella model.  The fit is conditioned on $X\geq 1$.  Using nonlinear least squares fitting with the constraint $b<0$, we have obtained near-identical values of the fitting parameters.  The results are shown in Table~\ref{tab:fitting}, and confirm yet again that the scaling behaviour of the rim-lamella model is `sandwiched' between the upper and lower bounds.  It can also be seen that the scaling behaviour of the rim-lamella model  is very close to the semi-empirical correlation~\eqref{eq:correlation}.
%
%
%\begin{table}
	%\centering
		%\begin{tabular}{|p{4cm}||p{1.5cm}|p{1.5cm}|p{1.5cm}|}
		%\hline
		 %& $a$ & $b$ & $p$ \\
		%\hline
		%\hline
		%Lower Bound & 1.00 & -- & 0 \\
		%\hline
		%{\textbf{RL Model, $\myRe = 10^4$}}     & 1.00 & -0.47 & -1.19\\
		%\hline
		%{\textbf{RL Model, $\myRe=10^6$ }}    &  0.99 & -0.47 & -1.17\\
		%\hline
		%{\textbf{RL Model, $\myRe=10^8$}}     & 1.00 & -0.47 & -1.14\\
		%\hline
		%Upper Bound & 1.00 & - 1.55 & -1 \\
				%\hline
				%%Correlation~\eqref{eq:correlation} & 1.00 & -0.34 & -1  \\
		%%\hline
		%\end{tabular}
		%\caption{Parameter values obtained from fitting the curve $Y=a+b X^p$  to the data emanating from the rim-lamella model. Advancing contact angle: $\thetadyn=\pi/2$.  `RL model' refers to the rim-lamella model, solved numerically (specifically, the numerical solution of the ODEs~\eqref{eq:RLregime1} and~\eqref{eq:RLregime2}).}
		%\label{tab:fitting}
%\end{table}
%
%
\begin{table}
	\centering
		\begin{tabular}{|p{4cm}||p{1.5cm}|p{1.5cm}|}
		\hline
		 & $a$ & $b$  \\
		\hline
		\hline
		Lower Bound & 1.00 & 0  \\
		\hline
		{\textbf{RL Model, $\myRe = 10^4$}}     & 1.01 & -0.45  \\
		\hline
		{\textbf{RL Model, $\myRe=10^6$ }}    &  1.01 & -0.46  \\
		\hline
		{\textbf{RL Model, $\myRe=10^8$}}     & 1.02 & -0.47  \\
		\hline
		Upper Bound & 1.00 & - 1.55 \\
				\hline
				%Correlation~\eqref{eq:correlation} & 1.00 & -0.34 & -1  \\
		%\hline
		\end{tabular}
		\caption{Parameter values obtained from fitting the curve $Y=a+b X^{-1}$  to the data emanating from the rim-lamella model. Advancing contact angle: $\thetadyn=\pi/2$.  `RL model' refers to the rim-lamella model, solved numerically (specifically, the numerical solution of the ODEs~\eqref{eq:RLregime1} and~\eqref{eq:RLregime2}).}
		\label{tab:fitting}
\end{table}
A plot showing the data, the bounds, and the model is given in Figure~\ref{fig:curve_fitting}.

\begin{figure}
	\centering
		\includegraphics[width=0.7\textwidth]{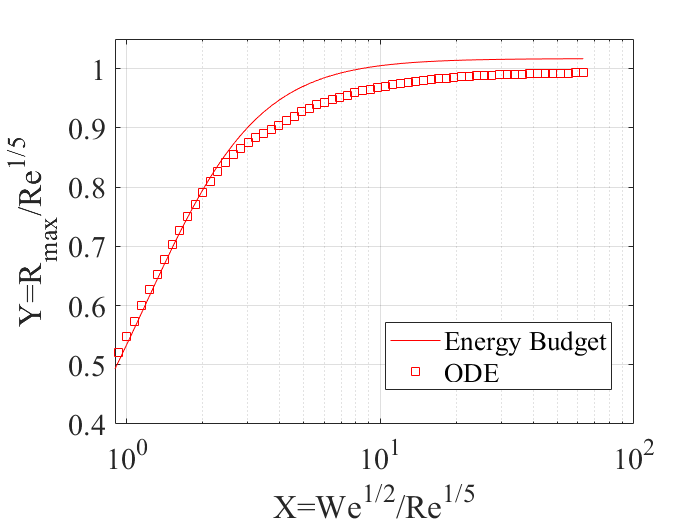}
		\caption{Comparison between energy-budget model and rim-lamella model, $\myRe=10^6$.}
	\label{fig:eb}
\end{figure}

Finally, we compare the results of the rim-lamella model with the energy-budget model due to Wildeman et al.~\cite{wildeman2016spreading}:
\begin{equation}
\underbrace{\frac{\alpha_w}{\myRe_D^{1/2}}D_m^2\sqrt{D_m-1}}_{\text{Viscous Dissipation}}+\underbrace{\frac{3}{\myWe_D}(1-\cos\thetadyn)D_m^2}_{\text{Surface energy at maximum spreading}}=1+\frac{12}{\myWe_D}-\phi.
\label{eq:wildeman}
\end{equation}
Here, $D_m=2R_{max}$ and $\myRe_D=2\myRe$ and $\myWe_D=2\myWe$; the dimensionless groups are based on diameters.
The energy budget is obtained by balancing the viscous dissipation and the surface energy at maximum spreading on the right-hand side with the initial kinetic energy of the droplet prior to impact (here normalized to one), the initial surface energy of the droplet prior to impact (here normalized to $12/\myWe_D$), and a `head loss' term.  The head loss accounts for internal flows which develop in the rim-lamella structure and which are otherwise not included in this simple balance.  The head loss is a simple fraction of the initial kinetic energy, thus $0\leq \phi<1$.   A comparison between rim-lamella model and energy-budget model is given  in Figure~\ref{fig:eb}, for the case $\myRe=10^6$.  Further comparisons are given in Table~\ref{tab:fitting1}.
We fix $\alpha_w=0.7$ and $\phi=0.5$ as given in the paper by Wildemen \textit{et al.}  There is excellent agreement between the two approaches, considering no effort has been made with fitting to bring about such agreement -- these are two independent methods, which provide near-identical prediction for the maximum spreading radius for $X\gg 1$ and $\myRe\gg 1$.  In particular, these results  confirm the necessity for including head loss in the energy-budget calculation.  
\begin{table}
	\centering
		\begin{tabular}{|p{4cm}||p{1.5cm}|p{1.5cm}|}
		\hline
		 & $a$ & $b$  \\
		\hline
		\hline
				Roisman~\cite{roisman2009inertia}       & 1.00 & - 0.37  \\
		\hline
		{\textbf{RL Model, $\myRe=10^6$ }}    &  1.01 & -0.46 \\
		
		\hline
    Wildeman, fit.~\cite{wildeman2016spreading}     & 1.00 & -0.65 \\
		\hline
		\end{tabular}
		\caption{Rigorous comparison between the rim-lamella model, the energy-budget model of Wildeman et al. fitted to the functional form $\rr= a\myRe^{1/5}-bX^{-1}$, and the correlation of Roisman \textit{et al.}, of a similar type.  Here, $X=\myWe^{1/2}/\myRe^{1/5}$.  Dimensionless groups $\myWe$ and $\myRe$ based on radius, not diameter.}
		\label{tab:fitting1}
\end{table}

Finally, we emphasize that one can glean insights from the rim-lamella model about the dynamics of the spreading (and not just the maximum spreading radius).  In this way, one can begin to think about what happens when more physical phenomena are introduced into the model, for instance a dynamic contact angle, surface heterogeneity, etc.
%
%\begin{figure}
	%\centering
		%\includegraphics[width=0.7\textwidth]{sample_plots}
		%\caption{Sample plots of the $R(t)$ for $\myRe=10^6$, $\myWe=10^3$, and various values of $\thetadyn$.  Note that the model loses physical relevance when $R\rightarrow 0$}.
	%\label{fig:sample_plots}
%\end{figure}
%
Such physical insights can help with practical design in applications, as stated earlier, in the Introduction.

\section{Conclusions}
\label{sec:conc}

Summarizing, we have revisited the problem of droplet impact and droplet spread on a smooth
surface in the case of a viscous fluid. This problem is well studied in the literature, and
there are two quite similar models which describe the maximum spreading radius, based on the dynamics of the rim-lamella structure~\cite{eggers2010drop,gordillo2019theory}.  We have revisited these models with three aims in mind:
\begin{enumerate}[noitemsep]
\item To develop a rim-lamella model rigorously from first principles, based on mass conservation of the rim-lamella structure, and Newton's law for same.
\item To conduct a mathematical analysis of the model, based on the theory of differential inequalities, and hence to establish the bounds~\eqref{eq:bounds}.
\item To use the rim-lamella model as an independent means of validating the energy-balance approach to predicting the maximum spreading radius, $\rr$.
\end{enumerate}
In fulfilling the first aim, we have shown that the model admits a singular solution.  We have investigated the regularization techniques applied in previous works to avert this singularity.  In those works, the regularization was implicit -- we make the regularization explicit here.  This enhances our understanding of such rim-lamella models. 

For the present purposes, we have specified a particular regularization technique which is implicit in Reference~\cite{gordillo2019theory}.  We  have analyzed the resulting ODE model using differential inequalties, thereby fulfilling our second aim.  In particular, we have established the bounds~\eqref{eq:bounds}, recalled here as:
\[
\myconst_1 \myRe^{1/5}-\myconst_2\sqrt{(1-\cos\thetadyn)} \myWe^{-1/2}\myRe^{2/5}\leq 
\frac{\rr}{R_0} \leq \myconst_1 \myRe^{1/5},
\]
The upper bound can be obtained from dimensional analysis, where as the lower bound can be obtained by reasoning out the effect of the surface tension on the spreading, as in Reference~\cite{roisman2009inertia}.  In the present work, we provide an alternative method of obtaining the bounds -- in particular, we obtain the lower bound from Gronwall's inequality applied to the rim-lamella model.

We have validated the bounds~\eqref{eq:bounds} with respect to the experimental literature on droplet spreading.  Our theoretical bounds are consistent with the experimental results on droplet spreading.  Furthermore, our numerical solutions of the rim-lamella model are in close agreement with the experiments, confirming the correctness of the approach and the validity of the chosen regularization technique.  Our results further validate the energy-budget approach to modelling the maximum spreading radius, and confirm the importance of including head loss in the latter.   The head loss accounts for internal flows which develop in the rim-lamella complex and which are otherwise not included in this simple balance.  In a previous computational study~\cite{wildeman2016spreading}, the head loss was taken to be one half the initial kinetic energy of the drop.  Our results based on the rim-lamella model are consistent with this assumption.  This fulfils our third aim.

Our approach yields a simple set of ordinary differential equations which can in principle be applied to a wide range of droplet-impact phenomena, beyond normal impact of droplets of Newtonian fluid on a uniform surface.   The model as developed here could be used as a basis for such applications.  Alternatively, it may be useful first to explore the singularity inherent in the rim-lamella models in more detail, and hence to formulate a model that is intrinsically well-posed, without having to resort at all to regularization techniques.

\subsection*{Acknowledgments}

Helpful discussions with Alidad Amirfazli, Yating Hu, and Chris Howland are acknowledged.

\end{document}